\documentclass[12pt]{iopart}   
\usepackage{iopams} 
\usepackage{graphicx}

\begin{document}

\title[Tipping time of a quantum rod]{Tipping time of a quantum rod}
\author{Onkar Parrikar}
\address{Birla Institute of Technology and Science - Pilani, Goa campus, Zuarinagar, Goa, India 403726}
\ead{onkarsp@gmail.com}
\begin{abstract}
The behaviour of a quantum rod, pivoted at its lower end on an impenetrable floor and restricted to moving in the vertical plane under the gravitational potential is studied analytically under the approximation that the rod is initially localized to a ``small-enough'' neighbourhood around the point of classical unstable equilibrium. It is shown that the rod evolves out of this neighbourhood. The time required for this to happen, i.e., the tipping time is calculated using the semi-classical path integral. It is shown that equilibrium is recovered in the classical limit, and that our calculations are consistent with the uncertainty principle. 
\end{abstract}
\pacs{03.65.-w, 03.65.Sq, 0.3.65.Xp}

\maketitle

\section{Introduction}
It is well known that a vertical classical rod at rest, pivoted at its lower end on an impenetrable floor will continue to remain vertical till an external perturbation moves it out of its state of unstable equilibrium. Therefore, if the system were to be isolated from any such disturbances, the rod would never tip off. The scenario changes dramatically when one replaces the classical rod in the above picture with a quantum rod. In the latter case, the uncertainty priciple causes the rod to transit out of equilibrium, i.e., tip over. Many texts on quantum mechanics, for instance \cite{sakurai}, ask the reader to estimate this tipping time using the uncertainty principle; but a na\"ive application can lead to incorrect results (see \cite{shegelski} for an interesting discussion). \\\\In this paper aimed at advanced undergraduates, graduate students and teachers in physics, we systematically analyze the evolution of a quantum rod out of unstable equilibrium. Our analysis involves the semi-classical approximation to path integral, and enables us to derive an analytical expression for the tipping time. It is also shown, that unstable equilibrium is recovered in the classical limit. The calculation is meant to explicitly illustrate and thus help students visualize, the drastic differences between classical and quantum mechanical evolution of systems.\\\\
The issue of quantum mechanical rolling out of unstable equilibrium was first addressed in the context of an inverted pendulum in \cite{cook}. A special case of the problem we discuss has been considered, in the light of W.K.B approximation in \cite{batista}\footnote{In \cite{batista}, the W.K.B approximation is used to calculate energy eigenfunctions and eigenvalues, and the time evolved state is calculated in the limit that expectation of energy of the rod is close to the potential energy at the point of unstable equilibrium.}. A variant of the problem has also been addressed before, where the centre of mass of the rod is localized within the base of support of the rod (in our case, the base is taken to be a point), and the tipping time is computed numerically \cite{shegelski,shegelski2}\footnote{In fact, the potential function used in \cite{shegelski,shegelski2} is of the form \(V_0\left(\cos(\theta-\theta_0)-\cos(\theta_0)\right)\), consisting of a minimum at \(\theta=0\) between two maxima. Thus the resulting situation is a barrier tunnelling problem.}. However, we seek an analytical expression for the tipping time.
\section{Formulation of the problem}
Consider a rigid rod of uniformly distributed mass `\(m\)' and negligible cross section, in contact with an impenetrable floor at the point \(O\). We assume that the rod is constrained to move in the vertical plane. Let \(C\) be the centre of mass of the rod, and let the distance \(OC\) be `\(a\)'. The moment of inertia \(I\) of the rod is given by \(I=\frac{4}{3}ma^2\). The potential under which the particle moves is the gravitational potential given by
\begin{eqnarray}
V(\theta)&= mga \cos\theta \qquad & \theta\in[-\pi/2,\pi/2]\nonumber\\&=\infty & \theta\notin[-\pi/2,\pi/2] \nonumber
\end{eqnarray}
where \(\theta\) is measured from the vertical position. The latter condition on \(V(\theta)\) follows from the fact that the floor is assumed to be impenetrable. The constant `\(g\)' is the acceleration due to gravity. \\\\
Since we need the rod to be localized around \(\theta=0\) initially, the wavefunction \(\psi(\theta,t)\) at \(t=0\) is chosen to be a Gaussian, with expectation of position \(<\theta>=0\) and expectation of canonical momentum \(<l>=0\), i.e.,
\begin{equation}
\psi(\theta,0)=\frac{1}{(\sqrt{\pi}\sigma \mathrm{Erf}(\pi/2\sigma))^{1/2}}\exp\left(-\frac{\theta^{2}}{2\sigma^{2}}\right)\label{1}
\end{equation}for \(\theta\in[-\pi/2,\pi/2]\) and zero elsewhere. Clearly, the wavefunction has the shortcoming that it is not continuous at \(\theta = \pm\pi/2\). The justification is that the parameter \(\sigma \ll 1\), in order for the initial state to be close to classical unstable equilibrium, and therefore \(\psi(\pm\pi/2,0)\approx0\). Note that with this ``approximate'' continuity, the wavefunction is normalized within the region \(\theta\in[-\pi/2,\pi/2]\). Also, we let \(\sigma\) be arbitrary (provided it is much less than one); as against \cite{cook,batista}, where \(\sigma\) is fixed by requiring that the expectation of \(H\) for \(\psi(\theta,0)\) is a minimum. \\\\
Finally, we are interested in calculating the time it takes for the rod to evolve out of the \(\sigma\)-neighbourhood, to which it is localized initially. We define this as the `tipping time'-\(t_{\mathrm{tip}}\) for the system. Note that this definition differs from the ones used before\cite{shegelski,batista,shegelski2}\footnote{While \cite{batista} defines tipping time as the time required for the rod to fall to the floor, \cite{shegelski,shegelski2} use the average time for which the center of mass is still inside the region to which it was localized initially.}. We identify the motion of the rod with the trajectory of the probability density maximum in the \(\theta\)-t space. This gives us a semi-classical understanding of the phenomenon of ``tipping'' of the quantum rod. The tipping time as defined here, is a measure of the time until which classical and quantum evolutions are close in some sense. Also, the definition helps us in generalising the conclusions of our calculation to all potentials (bounded from below) which have a point of unstable equilibrium. 

\section{Semiclassical Path integral}
The time evolution operator, i.e. the propagator is defined as \[G(\theta_2,t_2;\theta_1,t_1)=\left\langle\theta_2\left|\exp\left(-\frac{i(t_2-t_1)}{\hbar}H\right)\right|\theta_1\right\rangle\]
where \(H\) is the hamiltonian operator. If the eigenvalues \(E_n\) and eigenfunctions \(\phi_n(\theta)\) of \(H\) are known, then the propagator can be written as
\[G(\theta_2,t_2;\theta_1,t_1)= \sum_{n}\exp\left(-\frac{i(t_2-t_1)}{\hbar}E_n\right)\phi^*(\theta_2)\phi(\theta_1)\]
More importantly, the propagator is the Green function corresponding to the Schr\"odinger's time-dependent equation. Thus, knowledge of the propagator allows us to calculate the time evolution of any initial state under the hamiltonian \(H\). Note also, that normalizability of wavefunction at all times demands unitarity of the propagator.\\\\
There is another representation for the propagator called the path integral representation, in which the propagator can be written as \cite{feynman,kleinert}
\begin{equation}
G(\theta_{2},t_{2};\theta_{1},t_{1})=\int_{\theta(t_1)=\theta_1}^{\theta(t_2)=\theta_2} D[\theta(t)] \label{3} \exp\left(\frac{i}{\hbar}S[\theta(t)]\right)
\end{equation}
where \(S[\theta(t)]\) is the action corresponding to the Lagrangian \(L(\theta,\dot{\theta})\) along the path \(\theta(t)\), i.e. \[S[\theta(t)]=\int_{t_{1}}^{t_{2}}L(\theta,\dot{\theta})\rmd t\] The integration in (2) is over all the paths between \((\theta_1,t_1)\) to \((\theta_2,t_2)\); thus the term path integral. It can be shown, that for \(\hbar\rightarrow0\), the ``classical path'' which is an extremum for the action, provides the most dominant contribution to the path integral\cite{feynman,kleinert}; and that (2) can be written as
\begin{equation}
G(\theta_{2},t;\theta_{1},0)\approx G(0,t;0,0)\exp\left(\frac{i}{\hbar}S_{\mathrm{class}}\right)
\end{equation}
where \(S_{\mathrm{class}}\) is the action corresponding to the classical path, and \(G(0,t;0,0)\) is a factor independent of \(\theta_1\) and \(\theta_2\). In what follows, we will be using this ``semiclassical'' approximation to the path integral to calculate the propagator instead of using the Schr\"odinger's equation. The justification for this will be provided towards the end of this section.\\\\
Now, the fact that we need the initial state to be close to classical unstable equilibrium (\(\sigma\ll1\)) provides for some simplifying approximations. Observe that for \(t<t_{\mathrm{tip}}\), the wavepacket is largely contained within the \(\sigma\) neighbourhood of the point \(\theta=0\). This suggests that for calculating \(t_{\mathrm{tip}}\), one can practically replace the gravitational potential by \(V(\theta)\approx mga(1-\frac{\theta^{2}}{2})\)for \(-\pi/2\leq\theta\leq\pi/2\) and infinity elsewhere.\\\\
With the above simplification, the lagrangian for the system with \(-\pi/2\leq\theta\leq\pi/2\) becomes 
\[L(\theta,\dot{\theta})=\frac{1}{2}I\dot{\theta}^{2}+\frac{1}{2}ma^{2}\omega^{2}\theta^{2}\]
where \(\omega^{2}=g/a\). Without the restriction on configuration space, the corresponding hamiltonian would clearly be unbounded from below, and thus lead to non-normalizable eigenfunctions. But the hard floor provides for a bounded configuration space, and thus one expects to maintain normalizability.\\\\
The above lagrangian can be written in the familiar form 
\begin{equation}
L(\theta,\dot{\theta})=\frac{1}{2}Ma^2\dot{\theta}^{2}+\frac{1}{2}Ma^{2}\Omega^{2}\theta^{2} \label{2}
\end{equation}
with the redefined variables \(M=I/a^2\) and \(\Omega=(m/M)^{1/2}\omega\).
Notice that this is the Lagrangian for the linear simple harmonic oscillator (SHO) analytically continued to imaginary frequency. This is clearly a general feature of all potentials in a small neighbourhood around a point of unstable equilibrium; \(\omega^{2}\) being a measure of the curvature of the potential at that point.  But the system at hand differs from SHO in that our configuration space is bounded. Despite this, it will be shown that one can use an SHO-like propagator as the time evolution operator for the given system without violating unitarity for ``small enough'' times. \\\\
Coming back to the evaluation of the propagator, classical trajectories are expected to dominate the path integral as mentioned before. But because the configuration space is bounded in our case, there are multiple classical paths, owing to ``bouncing off'' from the boundaries. Therefore, an exact treatment would have to incorporate multi-instanton solutions  \cite{kleinert,holstein}. But if \(t\) is small enough, then the direct classical path- i.e. a monotonic solution of the classical equation of motion \[\ddot{\theta}(\tau)=\Omega^2\theta(\tau)\]with the boundary conditions \(\theta(0)=\theta_1\) and \(\theta(t)=\theta_2\)- will dominate in comparison to the other classical trajectories. This path can be easily obtained by imposing the above boundary conditions on the general solution \(c_1\sinh(\Omega\tau)+c_2\cosh(\Omega\tau)\), and is given by
\[\theta_{\mathrm{class}}(\tau)=\frac{\theta_{2}-\theta_{1}\cosh(\Omega t)}{\sinh(\Omega t)}\sinh(\Omega \tau)+\theta_{1}\cosh(\Omega \tau)\] The approximation clearly becomes better as \(t\rightarrow0\). Consequently, using (3), the propagator can be written down as 
\begin{eqnarray}
\fl G(\theta_{2},t;\theta_{1},0)\approx G(0,t;0,0)\nonumber \\
\exp\left[\frac{iMa^{2}\Omega}{2\hbar \sinh(\Omega t)}\left((\theta_{1}^{2}+\theta_{2}^{2})\cosh(\Omega t) - 2\theta_{1}\theta_{2}\right)\right] \label{4}
\end{eqnarray}where \[S_{\mathrm{class}}= \frac{Ma^{2}\Omega}{2\sinh(\Omega t)}\left((\theta_{1}^{2}+\theta_{2}^{2})\cosh(\Omega t) - 2\theta_{1}\theta_{2}\right)\]is the action\cite{feynman} corresponding to \(\theta_{\mathrm{class}}\) given above. A complete determination of the propagator however, would need determination of the position independent factor \(G(0,t;0,0)\). Moreover, we need to have a clear upper bound on \(t\) for which (\ref{4}) makes sense. We impose unitarity on the propagator for this purpose, and claim that (\ref{4}) is valid for those values of \(t\), for which \[\int_{-\pi/2}^{\pi/2}G^{*}(\theta,t;\theta',0)G(\theta',t;0,0)\rmd \theta'=\delta(\theta)\]is satisfied. But the left hand side of the above equation with the form of propagator suggested by (\ref{4}), is certainly not a delta function; the reason being that not all classical paths have been incorporated in (\ref{4}). Nevertheless it is plausible that one recovers a delta function- approximately for small enough \(t\), and exactly in the limit \(t \rightarrow0\). Explicitly, 
\begin{eqnarray}
\fl \int_{-\pi/2}^{\pi/2}G^{*}(\theta,t_;\theta',0)G(\theta',t;0,0)\rmd \theta'=\left|G(0,t;0,0)\right|^{2}\nonumber\\ \exp\left(-\frac{iMa^{2}\Omega\theta^{2}\cosh(\Omega t)}{2\hbar \sinh(\Omega t)}\right)\int_{-\pi/2}^{\pi/2}\exp\left(\frac{iMa^{2}\Omega\theta}{2\hbar \sinh(\Omega t)}\theta'\right)\rmd \theta'\nonumber\\
=\left|G(0,t;0,0)\right|^{2}\exp\left(-i\alpha\theta^{2}\cosh(\Omega t)\right)\frac{\sin(\alpha\pi\theta/2)}{\alpha\theta}\label{5}
\end{eqnarray}
where \(\alpha=Ma^{2}\Omega/(2\hbar \sinh(\Omega t))\). Notice that
\[\lim_{\beta\rightarrow\infty}\frac{\sin(\beta \theta)}{\theta}=\pi\delta(\theta)\]Therefore, if \(\alpha\rightarrow\infty\), the right hand side of (\ref{5}) approaches a delta function upto some multiplicative factors. Note that the phase \(\exp\left(-i\alpha\theta^2\cosh(\Omega t)\right)\) goes away as \(\theta\) is set to zero in the exponential owing to the delta function. Finally, to recover the unitarity condition, one can chose \(G(0,t;0,0)\) appropriately (up to a phase) so as to cancel out all the factors multiplying the delta function. Since only \(\left|G(0,t;0,0)\right|^{2}\) will appear in our calculation, the phase of \(G(0,t;0,0)\) does not make any difference. Therefore, following \cite{feynman}
\begin{equation}
G(0,t;0,0)\approx\left(\frac{Ma^{2}\Omega}{2\pi i\hbar \sinh(\Omega t)}\right)^{1/2} \label{6}
\end{equation}
In this sense, unitarity is respected (approximately) only for \(\alpha\gg1\). This gives us an upper bound for \(t\), i.e., \(\sinh(\Omega t)\ll Ma^2\Omega/2\hbar\), for which the above results are valid. We state all of this in Proposition 1 below-\\\\
\textbf{Proposition 1}: If \(t\) is such that \(\sinh(\Omega t)\ll Ma^2\Omega/2\hbar\), then the propagator \(G(\theta_{2},t;\theta_{1},0)\) is approximately given by 
\begin{eqnarray}
\fl G(\theta_{2},t;\theta_{1},0)\approx \left(\frac{Ma^{2}\Omega}{2\pi i\hbar \sinh(\Omega t)}\right)^{1/2}\nonumber\\
\exp\left[\frac{iMa^{2}\Omega}{2\hbar \sinh(\Omega t)}\left((\theta_{1}^{2}+\theta_{2}^{2})\cosh(\Omega t) - 2\theta_{1}\theta_{2}\right)\right]\label{7}
\end{eqnarray}
\Eref{7} becomes exact in the limit \(t\rightarrow0\).\\\\A few comments are in order at this point, about the use of semiclassical path integral as against the Schr\"odinger equation formalism. Observe that the semi-classical path integral allowed us to work with the approximate \(\theta^{2}\) potential in a small neighbourhood of the point of unstable equilibrium. This simplification is clearly not suited for finding the energy eigenfunctions using Schr\"odinger's equation. This is because in order to obtain the eigenfunctions, one has to impose Dirichlet boundary conditions on the solutions at \(\theta=\pm\pi/2\), where the \(\theta^2\) approximation clearly fails. Also, one could in principle solve for the eigenfunctions using the full form of the potential. The corresponding differential equation is called Mathieu 's equation (See \cite{batista} for an evaluation of the eigenfunctions in this manner). But the method is rather unwieldy for calculating the propagator. Thus, it is easier to use the semi-classical path integral, which also provides a more physical picture. On the other hand, our method has the limitation that one can only calculate small-time behaviour for systems which are well-localized initially. But as will be shown in the next section, this does not hamper the calculation of the tipping time for the rod.\\\\
In the calculations that follow, we will use the propagator derived above in calculating the time evolution of the quantum rod for \(t\) much less than the required upper bound . This will be used to determine the tipping time, and it will be shown that our results are valid if the quantum rod is initially ``localized enough'' about the point \(\theta=0\). 

\section{Evolution out of classical unstable equilibrium}
Proposition (1) enables us to calculate the evolution of the quantum rod whose state at time \(t=0\) is given by the wavefunction in (\ref{1}). The state of the system at a later time \(t\) is given by\[\psi(\theta,t)=\int_{-\pi/2}^{\pi/2}G(\theta,t;\theta',0)\psi(\theta',0)\rmd\theta'\]The unitarity of the propagator ensures normalization of the wavefunction at all times \(t\). For \(t\) satisfying the upper bound, one can use Proposition (1) in the above equation. 
\begin{eqnarray}
\fl \psi(\theta,t)\approx \frac{G(0,t;0,0)}{{(\sqrt{\pi}\sigma \mathrm{Erf}(\pi/2\sigma))^{1/2}}}\nonumber \\
\int_{-\pi/2}^{\pi/2}\rmd \theta'\exp\left[\frac{iMa^{2}\Omega}{2\hbar \sinh(\Omega t)}\left((\theta^{2}+\theta'^{2})\cosh(\Omega t) - 2\theta\theta'\right)-\frac{\theta'^{2}}{2\sigma^{2}}\right]\label{8}
\end{eqnarray}
Again, note that one is interested in the case \(\sigma \ll1\). Therefore, we use the approximation \(\mathrm{Erf}(\pi/2\sigma)\approx1\). Also, since \(\psi(\theta',0)\) drops down very rapidly for \(\left|\theta\right|>\sigma\), we can extend the domain of integration to \((-\infty,\infty)\). With this, the right hand side of (\ref{8}) becomes
\begin{eqnarray}
\fl \frac{G(0,t;0,0)}{{(\sqrt{\pi}\sigma)^{1/2}}}\exp\left(i\alpha\theta^{2}\cosh(\Omega t)\right)\nonumber\\
\int_{-\infty}^{\infty}\rmd \theta'\exp\left[(i\alpha\cosh(\Omega t)-\frac{1}{2\sigma^{2}})\theta'^{2}-2i\alpha\theta\theta'\right]\label{9}
\end{eqnarray}
which is clearly a Gaussian integral. The integral is well defined since the coefficient of \(\theta'^{2}\) term in the exponential has a negative real part, and is given by
\begin{eqnarray}
\fl \sqrt{\pi}\frac{G(0,t;0,0)}{{(\sqrt{\pi}\sigma)^{1/2}}}\exp\left(i\alpha\theta^{2}\cosh(\Omega t)\right) \nonumber\\
\frac{1}{(-i\alpha\cosh(\Omega t)+1/2\sigma^{2})^{1/2}}\exp\left[\frac{4\alpha^{2}\theta^{2}}{4i\alpha\cosh(\Omega t)-2/\sigma^{2}}\right]\nonumber
\end{eqnarray}
Having obtained \(\psi(\theta,t)\), one can now calculate the probability density \(P(\theta,t)=\left|\psi(\theta,t)\right|^{2}\). In doing so, we note that the phase factors in the above equation do not contribute. Thus, the probability density function is given by
\begin{eqnarray}
\fl P(\theta,t)=\frac{\sqrt{\pi}}{\sigma}\left|G(0,t;0,0)\right|^{2}\frac{1}{(\alpha^{2}\cosh^{2}(\Omega t)+1/4\sigma^{4})^{1/2}} \nonumber \\
\exp\left[-\frac{4\alpha^{2}\sigma^{2}\theta^{2}}{1+4\alpha^{2}\sigma^{4}\cosh^{2}(\Omega t)}\right] \label{10}
\end{eqnarray}

Remember that \(\alpha\) also has time dependence, i.e., \(\alpha=Ma^{2}\Omega/(2\hbar\sinh(\Omega t))\). The time dependence in (\ref{10}) can be made explicit
\begin{eqnarray} 
\fl P(\theta,t)=\frac{Ma^2\Omega\sigma}{\sqrt{\pi}(M^2a^4\Omega^2\sigma^{4}\cosh^{2}(\Omega t)+\hbar^2\sinh^{2}(\Omega t))^{1/2}} \nonumber \\
\exp\left[-\frac{M^2a^4\Omega^2\sigma^{2}\theta^{2}}{\hbar^2\sinh^{2}(\Omega t)+M^2a^4\Omega^2\sigma^{4}\cosh^{2}(\Omega t)}\right] \label{11}
\end{eqnarray}
At \(\theta=0\), the probability density is a monotonically decreasing function of time. For \(\theta\neq0\) the function \(P(\theta,t)\)  increases with time, from its value at \(t=0\) till it reaches a maximum, and then decays off to zero as \(t\rightarrow\infty\) (\Fref{1}). The path traced by this maximum of probability density in the \(\theta-t\) space, can be interpreted as the classical motion of the centre of mass of the rod. Moreover, if \(\theta_{1}<\theta_{2}\), then the time at which the probability density maximum occurs at \(\theta_{1}\) is less than the time at which it occurs at \(\theta_{2}\). 
\begin{figure}[htbp]
\centering
\includegraphics[height=10cm, width=10cm]{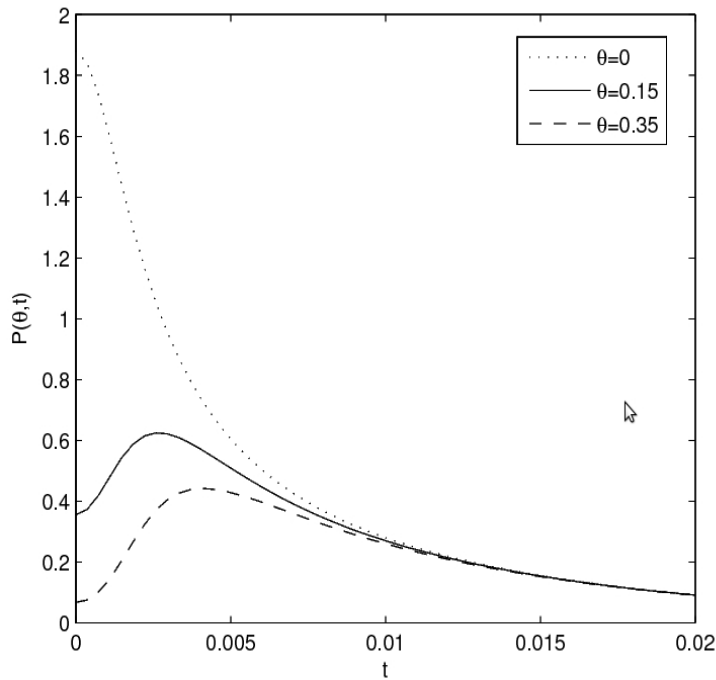}
	\caption{\(P(\theta,t)\) v/s \(t\) for different values of \(\theta\), with \(\sigma=0.3\) radians, \(\Omega=100\) s\(^{-1}\) and \(M\sim10^{-28}\) kg.}
	\label{1}
\end{figure}
Thus, the probability density maximum moves away from \(\theta=0\), which is what we interpret as the tipping of the rod. The tipping time \(t_{\mathrm{tip}}\) from our previous discussion, can be defined as the time at which the maximum occurs at \(\theta=\pm\sigma\). Mathematically, \(t_{\mathrm{tip}}\) is given by
\begin{equation}
\left(\frac{dP(\sigma,t)}{dt}\right)_{t=t_{\mathrm{tip}}}=0
\end{equation}
From (\ref{11}) and the definition above, \(t_{\mathrm{tip}}\) for the quantum rod is found to be
\begin{equation}
t_{\mathrm{tip}}=\frac{1}{\Omega}\sinh^{-1}\left(\frac{Ma^{2}\Omega\sigma^{2}}{(\hbar^{2}+M^{2}a^{4}\Omega^{2}\sigma^{4})^{1/2}}\right)\label{13}
\end{equation}
Note that since \(\sigma\ll1\), \(\sinh(\Omega t_{\mathrm{tip}})\ll Ma^{2}\Omega/2\hbar\). Therefore, the hypothesis of Proposition (1) has been shown to be true, till the rod tips. This completes the justification for using the propagator of Proposition (1) in the above calculation. \\\\
We can now switch back to the original variables \(m, I\) and \(\omega\). In terms of these, 
\begin{equation}
t_{\mathrm{tip}}=\frac{\sqrt{\kappa}}{\omega}\sinh^{-1}\left(\frac{\sqrt{\kappa}ma^2\omega\sigma^2}{(\hbar^{2}+\kappa m^2a^{4}\omega^{2}\sigma^{4})^{1/2}}\right)\label{14}
\end{equation}
where \(\kappa=I/ma^2=\frac{4}{3}\). Notice that \(\hbar/ma^2\omega\sim\lambda_{\mathrm{dB}}/a\), where \(\lambda_{\mathrm{dB}}\) is the reduced de Broglie wavelength for the rod \cite{cook}. In the quantum domain, \(\lambda_{\mathrm{dB}}>a\). Therefore the \(\sigma^{2}\) term can be dropped in comparison to the other term in the denominator of (\ref{14}). This gives \begin{equation}t_{\mathrm{tip}}\approx\frac{\sqrt{\kappa}}{\omega}\sinh^{-1}\left(\frac{\sqrt{\kappa}ma^{2}\omega\sigma^{2}}{\hbar}\right)\label{15}\end{equation}
The approximately quadratic dependence on \(\sigma\) suggests that, more localized is the initial state, the faster it will tip off. It can also be checked that \(t_{\mathrm{tip}}\) is a monotonically decreasing function of \(\omega\); reminiscent of the fact that stronger is the gravitational field, faster is the tipping. On the other hand \(\lambda_{\mathrm{dB}}\ll a\) represents the classical domain. Observe from (\ref{11}), that in the limit \(\hbar\rightarrow0\) and \(\sigma\rightarrow0\), \(P(\theta,t)\rightarrow0\) for \(\theta\neq0\). Therefore, the rod remains localized at the point \(\theta=0\); ergo classical equilibrium is recovered.\\\\
It is worthwhile noticing that the above analysis would go through for any potential \(U(\theta)\) bounded from below, in the neighbourhood of a point of unstable equilibrium \(\theta_0\), with \(\omega^2=\frac{1}{ma^2}\left|U''(\theta_0)\right|\). Thus, provided \(\sigma\ll1\), (\ref{14}) and (\ref{15}) are generically true for all potentials outlined above, with slowly varying curvatures. One can also understand (\ref{15}), at least crudely, in terms of the uncertainty principle. Note that because \(\sinh(x)\sim x\) for \(x\ll1\), one can linearize (\ref{15}) as
\[t_{\mathrm{tip}}\approx\frac{I\sigma^{2}}{\hbar}\]Since \(\sigma\sim\Delta\theta\), the uncertainty in angular position for the initial wavefunction, and \(I(\sigma/t_{\mathrm{tip}})\sim\Delta l\), the uncertainty in angular momentum, the above equation takes the form
\[\Delta\theta.\Delta l \approx \hbar\]
which is in accordance with the uncertainty principle. Therefore, the tipping of the quantum rod can be understood as having been triggered by the uncertainty in angular momentum engendered by localisation of the initial state. 
\section{Summary}
The quantum evolution of a rod out of unstable equilibrium was analyzed using the semiclassical path integral. It was shown that for small enough times, the propagator for the system is the same as that of a simple harmonic oscillator, analytically continued to imaginary frequency. This was used to compute the tipping time of the quantum rod. In the quantum domain, it was shown that the tipping time has an approximately quadratic dependence on localisation. It was also noticed that in the limit \(\hbar\rightarrow0\), the classical equilibrium can be recovered, and that ``tipping'' in this sense is purely a quantum mechanical phenomenon. 
\section{Acknowledgments}
I thank Prof S Ramaswamy for his valuable guidance and discussions and Ashok Ajoy for introducing me to this problem. I also thank the referee for his valuable suggestions towards improvement of the manuscript. 
\section*{References}
\bibliographystyle{unsrt}

\end{document}